\documentclass[10pt]{article}
\usepackage[preprint]{tmlr}
\usepackage{amsmath,amssymb,amsthm}
\usepackage{booktabs}
\usepackage{graphicx}
\usepackage{hyperref}
\usepackage{url}

\theoremstyle{plain}
\newtheorem{theorem}{Theorem}
\newtheorem{proposition}{Proposition}
\newtheorem{corollary}{Corollary}
\theoremstyle{definition}
\newtheorem{assumption}{Assumption}

\title{A Closed-Form Estimator and Diagnostic Battery for\\
Anchor--Judge Error Correlation, Under a Single-Common-Factor Model}

\author{\name Veerendra Kumar Sunkavalli \email veeru.svk@gmail.com \\
      \addr Independent Researcher}

\def\month{08}
\def\year{2026}
\def\openreview{\url{https://openreview.net/forum?id=XXXX}}

\makeatletter
\providecommand{\doi}{}
\renewcommand{\doi}[1]{\href{https://doi.org/#1}{doi:\nolinkurl{#1}}}
\makeatother

\begin{document}
\maketitle

\begin{abstract}
When an external reference set (an anchor) is used to decompose an LLM-judge panel's error into a
quality signal and a shared common-mode error, standard practice assumes the anchor is
uncontaminated: its error uncorrelated with the judges' shared error. We study when that assumption
can be dropped and replaced by an estimate. Under a single-common-factor model, $\ge2$ judges and
$\ge2$ anchors point-identify the quality variance, the common-mode variance, and each anchor's
contamination correlation $\rho_k$ in closed form, with an exact per-anchor-pair failure
boundary; a designated
clean-anchor estimator, by contrast, reports a contaminated companion anchor as fully clean once
its trusted anchor is itself contaminated. Because the single-common-factor assumption is itself
untestable, the estimator ships gated behind a calibrated diagnostic battery (judge-covariance
dispersion; over-identification; a family-block test from judge metadata, with a family-blocked
estimator that removes family-level shared-residual bias exactly), bootstrap confidence intervals
with measured coverage, and a weak-identification screen. A proposition maps which violations
bias $\rho_k$, in which direction, and which evade detection. For ordinal scores we show an
identification hierarchy: with all variables ordinal, $\rho_k$ is not identified at any number of
anchors; with ordinal judges and $\ge3$ continuous anchors it is, and we give an estimator for
that case. On real data the validation is asymmetric, and we say so plainly: the diagnostics are
validated in the rejecting direction (both real panels we test, human raters and a six-provider
LLM panel, are correctly rejected by the model-adequacy pre-test, Test~A; specificity on a real
\emph{adequate} panel is untested, none being available), while the estimator is validated in simulation and stress-tested semi-synthetically under oracle
calibration (a robustness observation, not a clean-model validation; Section~7); no real panel has
yet passed that adequacy pre-test, so the estimator has not been
validly applied to a real panel; the pre-test exists precisely to say so. All results replay
offline from shipped, checksummed artifacts; the pre-registration and its amendments are included.
Relative to MTMM method-factor models, identifiability is not new; the closed form, the exact
boundary, the violation taxonomy, and the calibrated battery for the LLM-judge setting are the
contribution.
\end{abstract}

\section{Introduction}
LLM-as-a-judge panels score model outputs at scale, and a recurring worry is that judges make
\emph{correlated} errors from shared pretraining, style preferences, or prompt framing, so averaging
does not cancel the error; measured directly, a nine-judge panel can carry only about two
independent votes' worth of information \citep{kohli2026nine}. A common remedy anchors the judges to an external reference set and
estimates the shared error against it \citep{care2026}. Existing anchored decompositions treat the anchor as
\emph{clean}: its error is assumed uncorrelated with the judges' shared error. If the anchor's
builders share the judges' biases, the decomposition is silently wrong, with no check.

This paper asks when that clean-anchor assumption can be replaced by a measurement of the
anchor's contamination $\rho$. One status statement up front: the diagnostics we ship are
validated on real panels; the estimator is not yet, because every real panel we tested fails the
model pre-test. Its current status is a correct tool awaiting a qualifying panel, and Section~8
states what qualifying requires. Identifiability of a
free-loading method factor from multiple indicators is classical (Section~\ref{sec:related}); we do
not claim it. What we contribute, \emph{under an explicitly stated single-common-factor model}, is a
closed-form estimator specialised to the judge/anchor structure; an exact characterisation of when
it fails; a proposition delimiting which model violations bias the estimate; and, because
the single-common-factor assumption is itself untestable, a calibrated diagnostic battery that
detects the violations that matter.

\paragraph{Contributions.}
\begin{enumerate}
\item \textbf{Closed-form estimator (Theorem~\ref{thm:id}).} Under the single-common-factor model,
$(\sigma_t^2,\sigma_c^2,\{\rho_k\})$ from a $\ge2$-judge panel and $\ge2$ anchors, with no
clean-anchor and no anchor-difference assumption. Identifiability itself is inherited from MTMM; the
closed form and the identifying role of the judge panel are what we add.
\item \textbf{Exact non-identification boundary (Corollary~\ref{cor:boundary}).} A measure-zero
condition $\beta_k=\sigma_c^2$, with a characterised positive-measure weak-identification
neighbourhood, the practically relevant failure.
\item \textbf{Which violations matter (Proposition~\ref{prop:oe}).} A second common factor loading
uniformly across judges and anchors is observationally equivalent to inflating $\sigma_t^2$ and
leaves $\rho_k$ \emph{unbiased}; only \emph{asymmetric} loadings bias $\rho_k$
(Proposition~\ref{prop:oe}); the harmful shared-residual case is treated in contribution~4.
\item \textbf{A calibrated three-test diagnostic battery.} A judge-covariance dispersion test
($\ge3$ judges; non-uniform judge loadings), a $\ge3$-anchor over-identification test (non-uniform
anchor loadings), and a family-block test (Test~C: shared family-level judge residuals, from judge
metadata alone, at any anchor count), each with a null-calibrated $5\%$ false-positive threshold;
the first two with reported power curves. A family-blocked variant of the estimator removes the
shared-residual bias exactly when the panel spans $\ge2$ families.
\item \textbf{Inference:} item-bootstrap confidence intervals with measured coverage at the
nominal level, and a weak-identification screen (a studentized denominator, in the spirit of
weak-instrument first-stage diagnostics) that flags boundary proximity from the data alone.
\item \textbf{An ordinal identification hierarchy (Section~\ref{sec:ordinal}):} with all variables
ordinal, $\rho_k$ is not identified at any number of anchors; with ordinal judges and $\ge3$
continuous-scored anchors it is, and we give a working estimator for that case.
\item \textbf{Positioning against standard estimation:} on identical data the closed form and
full-information ML (fitted with a standard SEM package) are statistically indistinguishable; the
closed form's value is transparency and the exact boundary diagnosis, not efficiency.
\item \textbf{An honest failure map:} finite-sample behaviour, weak identification near the
boundary, and a finite-sample upward bias of the mixed-scale estimator near $\rho=0$ at small $N$.
\end{enumerate}

\section{Related work and positioning}
\label{sec:related}
\paragraph{The identifiability is not new; we state the difference.} Our model
($J_j=t+c+e_j$, $A_k=t+\lambda_k c+u_k$) is a two-factor model with a shared common-mode/method
factor and correlated errors across methods; point-identifying factor variances and cross-loadings
from $\ge2$ indicators per factor is standard in multitrait--multimethod (MTMM) common-method-factor
models with correlated errors (for the model family, its identifiability, and the critique of its variants, see
\citealp{campbell1959,lance2002,bollen1989,anderson1956}), and the one-clean-anchor special case closely
parallels the reference-method CT-C$(M{-}1)$ design \citep{eid2000,eid2003}. Relative to that literature
our additions are: (a) a closed-form estimator for the unit-loading judge parameterisation; (b) the
exact non-identification boundary; (c) Proposition~\ref{prop:oe} on which violations bias $\rho_k$;
and (d) the calibrated diagnostic battery for the judge/anchor setting. Recent LLM-judge aggregation
work models quality plus shared confounders and recovers quality without ground truth; it uses any
human anchor only as a trusted labelling device and has no parameter for anchor--judge error
correlation \citep{care2026}, and its omitted-confounder bias result overlaps our misspecification
analysis, which we cite \citep{care2026} rather than claim. The diagnostic-testing literature on estimating error rates without a gold standard
\citep{hui1980,dawidskene1979} and on relaxing conditional independence between tests
\citep{dendukuri2001}, whose random-effects dependence models play the role our
shared-residual analysis plays here, estimated there by Bayesian machinery with informative priors
where we give the closed-form bias expression and a metadata-based test, established that such
correlations can be weakly identified and
model-dependent \citep{albertdodd2004}; our boundary and weak-identification results are the
continuous-score, judge-panel analogue, and we credit that literature for the phenomenon. A
contaminated anchor is an invalid instrument; our two-anchor system is an over-identified moment
system \citep{conley2012}.

\section{Model and identification}
\label{sec:model}
\begin{assumption}[Single common factor]\label{ass:scf}
All variables are mean-zero. There is one latent quality $t$ ($\mathrm{Var}=\sigma_t^2$) and one
common-mode factor $c$ ($\mathrm{Var}=\sigma_c^2$), $t\perp c$. Judges and anchors are
$J_j=t+c+e_j$ ($j=1,\dots,p$) and $A_k=t+\tfrac{\beta_k}{\sigma_c^2}c+u_k$ ($k=1,\dots,m$), with
$e_j,u_k$ mutually independent and independent of $(t,c)$. This is the load-bearing and untestable
assumption; Sections~\ref{sec:diag}--\ref{sec:limits} study its violation.
\end{assumption}

\begin{assumption}[Conditional judge-residual independence]\label{ass:jri}
The judge residuals $e_j$ are mutually uncorrelated: all shared judge variation is carried by the
single common-mode factor $c$. This names, as its own assumption, the judge-residual component
already contained in Assumption~\ref{ass:scf}'s mutual-independence clause; we state it separately
because it is the empirically fragile part and the one whose violation this paper studies:
LLM judges built on a shared base model, prompt template, or preference-tuning lineage plausibly
share residual correlation \emph{beyond} $c$, violating it. A \emph{homogeneous} such residual
(loading uniformly across judges) is misattributed by the estimator. The exact bias follows from
Theorem~\ref{thm:id} with $K$ shifted by the residual variance: with at least one clean anchor
the effect is pure attenuation, $|\hat\rho_k|<|\rho_k|$ (a false ``clean''); with both anchors
contaminated the direction is configuration-dependent: attenuation in the moderate-contamination
regime (simulation: a uniform judge residual of strength $0.6$ drives $\hat\rho_2$ down,
$0.701\to0.565$, while inflating $\hat\sigma_c^2$, $0.80\to1.13$), but inflation of
$\hat\rho_k$ itself, even past $|\rho|=1$, near the identification boundary. Because it is uniform across judges, the dispersion test (Test~A) does \emph{not} detect
it; it is the judge-side analogue of the asymmetric second factor of Proposition~\ref{prop:oe} and
is visible in principle to the over-identification test (Test~B) when the anchors are unaffected,
but only weakly at realistic scale: measured power at residual strength $0.6$ is $0.06$ at
$N{=}500$ (its own false-positive rate), $0.16$ at $N{=}2000$, and $0.99$ only at $N{=}10^4$
(Section~5), so in practice the family-blocked machinery of Section~5 is the operative remedy. We
foreground this as the empirically dominant failure mode in LLM evaluation.
Section~\ref{sec:diag} gives a partial remedy: when the residual is family-level and the panel
spans $\ge2$ families, a family-blocked estimator removes the bias and a third test detects the
residual from judge metadata alone; the panel-wide case remains open.
\end{assumption}
Here $\beta_k:=\mathrm{Cov}(A_k,c)$ and the contamination is
$\rho_k:=\beta_k/(\sigma_{a_k}\sigma_c)$, $\sigma_{a_k}^2$ the total error variance of anchor $k$;
$\rho_k=0$ is the clean anchor. The unit loadings on the judges fix the scale of $t$. Observable
second moments: $K=\mathrm{Cov}(J_i,J_j)_{i\ne j}=\sigma_t^2+\sigma_c^2$;
$M_k=\mathrm{Cov}(\bar J,A_k)=\sigma_t^2+\beta_k$;
$P_{k\ell}=\mathrm{Cov}(A_k,A_\ell)=\sigma_t^2+\beta_k\beta_\ell/\sigma_c^2$;
$\mathrm{Var}(A_k)=\sigma_t^2+\sigma_{a_k}^2$.

\begin{theorem}[Closed-form identification under Assumptions~\ref{ass:scf} and~\ref{ass:jri}]
\label{thm:id}
Under Assumptions~\ref{ass:scf} and~\ref{ass:jri} (the latter contained in the former's
independence clause, and doing the identifying work in $K=\sigma_t^2+\sigma_c^2$), with $p\ge2$
judges and $m\ge2$ anchors, whenever $(K+P_{12})-(M_1+M_2)\ne0$,
\begin{equation}
\sigma_t^2=\frac{KP_{12}-M_1M_2}{(K+P_{12})-(M_1+M_2)},\label{eq:st2}
\end{equation}
and $\sigma_c^2=K-\sigma_t^2$, $\beta_k=M_k-\sigma_t^2$,
$\sigma_{a_k}^2=\mathrm{Var}(A_k)-\sigma_t^2$, $\rho_k=\beta_k/(\sigma_{a_k}\sigma_c)$.
No clean-anchor and no anchor-difference assumption is required; with $p<2$ the quantity $K$ is
undefined and the parameters are not identified.
\end{theorem}
\begin{corollary}[Exact boundary and weak-identification neighbourhood]\label{cor:boundary}
The denominator of \eqref{eq:st2} equals $(\sigma_c^2-\beta_1)(\sigma_c^2-\beta_2)/\sigma_c^2$;
identification through this anchor pair fails exactly at $\beta_k=\sigma_c^2$ (measure zero);
with $m>2$ anchors the closed form averages over pairs, and identification of the remaining
parameters fails only when \emph{every} anchor sits at the boundary, since any non-degenerate pair
recovers $\sigma_t^2$ and hence each $\rho_k$. In a positive-measure
neighbourhood the estimator is weakly identified and its variance grows (Table~\ref{tab:boundary}).
\end{corollary}

\begin{proposition}[Which violations bias $\rho_k$]\label{prop:oe}
Add a second common factor $d\perp(t,c)$ with variance $\sigma_d^2$, loading $g$ on every judge and
$h$ on every anchor. Then, for $m=2$, there exists a single-common-factor model reproducing the observable
moments $(K,M_k,P_{12})$ exactly (for $m\ge3$ an asymmetric second factor is generically \emph{not}
reproducible by any single-factor model, which is exactly why the over-identification test can
detect it); when $g=h$ the equivalent parameters satisfy $\beta_k'=\beta_k$,
$\sigma_c^{2\prime}=\sigma_c^2$, and $\sigma_t^{2\prime}=\sigma_t^2+g^2\sigma_d^2$. Hence a uniformly
loaded second factor leaves $\rho_k$ unbiased (it is absorbed into the quality variance), and
$\rho_k$ is biased only to the extent the second factor loads \emph{asymmetrically} across judges
versus anchors. The $g=h$ case is, in effect, definitional: a factor loading identically on every
judge and anchor is indistinguishable from the unit-loading quality signal $t$, so ``harmless'' here
means ``reparameterises $t$'', not ``detected and corrected''. A genuinely shared bias of exactly
this symmetric shape would be silently credited as quality, and $\hat\sigma_t^2$ is inflated by
$g^2\sigma_d^2$ with no warning: users of the quality variance itself (e.g.\ for signal-to-noise
assessments) inherit that bias even though $\rho_k$ does not.
\end{proposition}
Proof (SymPy-verified) in Appendix~\ref{app:proof}; the bias-vs-asymmetry relationship is confirmed
in simulation (Section~\ref{sec:diag}). Estimates with $\sigma_c^2\le0$ or $\sigma_t^2\notin(0,K)$
are out-of-range (Heywood-type) and reported as such, not clipped. The closed form is not
range-restricted: $|\hat\rho_k|>1$ is possible and is itself evidence of misspecification or weak
identification (the raw real panel of Section~7 produces $1.026$); any use of the estimator should
be gated behind the adequacy pre-test (Test~A) and the battery.

\section{Simulation study}
\label{sec:sim}
Deterministic, seeds $\{11,\dots,88\}$ (one run per seed); each entry is mean$\pm$\emph{standard
deviation} across the eight seeds, emitted to
\texttt{p4\_core\_results.json}, \texttt{p4\_extended\_results.json}, \texttt{p4\_v2\_results.json},
\texttt{p4\_opchar\_results.json}. Quantities are \emph{simulated} unless marked [analytic]; these
are estimator-correctness and finite-sample characterisations, not evidence about real judges.

\paragraph{Recovery under contamination ($N{=}40{,}000$).}
With neither anchor clean and anchor reliabilities unknown, recovery is within $0.02$, including
statistically identical anchors: true $(\rho_1,\rho_2)=(0.3,0.7)\!\to\!(0.299\pm0.017,0.701\pm0.007)$;
$(0.1,0.9)\!\to\!(0.096\pm0.020,0.900\pm0.003)$; $(0.5,0.5)\!\to\!(0.500\pm0.012,0.502\pm0.009)$;
$(0.2,0.8)\!\to\!(0.198\pm0.019,0.801\pm0.005)$. This confirms the closed form inverts the moments;
it is not evidence the model holds for real judges.

\paragraph{Finite sample and out-of-range.}
Recovery of $\rho_2{=}0.7$: $0.718\pm0.067$ ($N{=}100$), $0.680\pm0.055$ ($500$),
$0.690\pm0.024$ ($2000$), $0.701\pm0.007$ ($40000$); no out-of-range estimates arose here. At small
$N$ the estimator is high-variance and should be used with interval estimates.

\paragraph{Weak identification near the boundary (Table~\ref{tab:boundary}).}
As $\beta_1$ approaches $\sigma_c^2$ the denominator of Theorem~1 shrinks: point estimates remain
approximately unbiased through $\rho_1=0.98$, and at $\rho_1=0.99$ the standard deviation grows
fifteen-fold ($0.955\pm0.061$), the practical signature of the measure-zero boundary. The
weak-identification screen of Section~\ref{sec:inference} is designed to flag exactly this regime.
\begin{table}[t]\centering
\caption{Approaching the boundary ($\sigma_c^2=0.8$, $\sigma_{a_1}=0.9$).}\label{tab:boundary}
\begin{tabular}{ccc}
\toprule $\rho_1$&$\beta_1$&$\hat\rho_1$ (sd)\\
\midrule
0.90&0.724&$0.900\pm0.004$\\
0.95&0.765&$0.949\pm0.003$\\
0.98&0.789&$0.979\pm0.004$\\
0.99&0.797&$0.955\pm0.061$ (variance blow-up)\\
\bottomrule\end{tabular}\end{table}

\section{Misspecification diagnostic battery}
\label{sec:diag}
By Proposition~\ref{prop:oe}, only \emph{asymmetric} second factors bias $\rho_k$. Table~\ref{tab:bias}
confirms it: holding judge loading $g=0.6$ and varying anchor loading $h$, the bias in $\hat\rho_2$
is $0.001$ at $h=g$; away from $h=g$ it is non-monotone in $|g-h|$ and changes sign at large
asymmetry (Table~\ref{tab:asym}: $\hat\rho_2$ rises $0.761\to0.780$, dips to $0.769$, then
crosses to $0.565$ against true $0.7$), so asymmetry determines \emph{that} the estimate is
biased, not the direction or size. We therefore target asymmetry with two complementary,
null-calibrated tests.

\begin{table}[t]\centering
\caption{Bias in $\hat\rho_2$ (true $0.7$) vs.\ loading asymmetry $|g-h|$ (judge $g{=}0.6$).}\label{tab:asym}
\label{tab:bias}
\begin{tabular}{cccccc}
\toprule $h$ & 0.6 & 0.5 & 0.4 & 0.3 & 0.0\\
$|g-h|$ & 0.00 & 0.10 & 0.20 & 0.30 & 0.60\\
$\hat\rho_2$ & 0.701 & 0.761 & 0.780 & 0.769 & 0.565\\
\bottomrule\end{tabular}\end{table}

\paragraph{Test A: judge-covariance dispersion (needs $p\ge3$).}
Under Assumption~\ref{ass:scf} all off-diagonal judge covariances are equal; a second factor with
non-uniform judge loadings makes them unequal. The statistic is the coefficient of variation of the
off-diagonal judge covariances. It requires $\ge3$ judges (with $p{=}2$ there is a single
off-diagonal), a caveat we flag since identification itself needs only $p\ge2$.

\paragraph{Test B: $\ge3$-anchor over-identification.}
With $m\ge3$, $\sigma_t^2$ is estimated from each anchor pair; under Assumption~\ref{ass:scf} all
agree. Non-uniform \emph{anchor} loadings break the agreement; the statistic is the spread of
$\hat\sigma_t^2$ across pairs, and it fires precisely where Test~A is blind.

\paragraph{Operating characteristics (Table~\ref{tab:opchar}).}
Thresholds are set at the $95$th percentile of each statistic's null distribution (correctly
specified model, $1000$ seeds), giving a $5\%$ false-positive rate by construction at every $N$
tested; this is a correct-model, Gaussian property (heavy tails inflate it to $0.083$, Section~6,
and specificity on real adequate panels is untested). Power (probability of exceeding threshold) rises with both signal strength and $N$, from a floor
that is genuinely uninformative: at spread $0.2$ Test~A's power is $0.04$ at $N{=}500$ and $0.11$
at $N{=}2000$, at or below its own false-positive rate, so mild violations at realistic $N$ are
invisible to it. Test~B's
power is measured against its advertised target: a second factor shared by $\ge2$ anchors with
non-uniform loadings ($h=(s,\,s/2,\,0)$ at strength $s$); a factor loading on a \emph{single}
anchor is not a violation at all (it is absorbed into that anchor's idiosyncratic variance, and
measured power equals the false-positive rate there).
\begin{table}[t]\centering
\caption{Battery operating characteristics. FP = false-positive rate under the correct model;
power at second-factor strength $0.4$ / $0.6$.}\label{tab:opchar}
\begin{tabular}{lccc}
\toprule
$N$ & FP (A, B) & Test A power (0.4/0.6) & Test B power (0.4/0.6)\\
\midrule
500   & 0.05 / 0.05 & 0.23 / 0.91 & 0.50 / 0.99\\
2000  & 0.05 / 0.05 & 0.93 / 1.00 & 0.99 / 1.00\\
10000 & 0.05 / 0.05 & 1.00 / 1.00 & 1.00 / 1.00\\
\bottomrule\end{tabular}\end{table}

\paragraph{Two blind spots, stated plainly.}
First: by Proposition~\ref{prop:oe} a \emph{uniformly} loaded second factor ($g=h$) is invisible to
both tests and to any second-moment statistic, but it does not bias $\rho_k$; there is nothing to
correct when it occurs. Second, and harmful: a \emph{shared judge residual}
(Assumption~\ref{ass:jri}) biases $\hat\rho_k$ (attenuation toward a false ``clean'' in the
typical regime; inflation near the boundary), Test~A cannot see it, and Test~B responds only
weakly below $N\approx10^4$ (judge-side power $0.06$/$0.16$/$0.99$ at $N{=}500$/$2000$/$10^4$,
strength $0.6$, versus $0.99$ at $N{=}2000$ for its anchor-side target; Table~\ref{tab:opchar}). The family-blocked estimator below removes
this bias exactly when the residual is family-level and the panel spans $\ge2$ families; what
survives is the \emph{panel-wide} residual (a shared prompt template affecting every judge), which
no within-panel statistic can separate from the common mode. Practitioners with single-family
panels should treat small $\hat\rho_k$ as unverified, not as clearance; and a panel generated
under a shared prompting protocol or template induces exactly this undetectable panel-wide case,
so on such panels the estimator should not be used at all, whatever the diagnostics say.

\paragraph{Family-blocked estimator, and Test C.} Judge lineage is observable metadata. Model the
family-level residual explicitly, $J_j=t+c+f_{b(j)}+e_j$ with $f_b$ independent across families:
then cross-family judge covariances equal $\sigma_t^2+\sigma_c^2$ exactly (the residual cancels),
so computing $K$ from \emph{cross-family pairs only} restores the closed form unchanged [analytic;
SymPy-verified], while the within-minus-cross gap estimates the family-residual variance directly,
$K_{\text{within}}-K_{\text{cross}}=\sigma_f^2$, giving a third diagnostic (Test~C) that needs
$\ge2$ families with at least one family containing $\ge2$ judges (so $p\ge3$); no extra anchors,
no over-identification machinery. Singleton families' residuals enter only through the cross terms. The
cross-family covariance contains no family terms at all, so the blocked estimator's exactness does
not depend on family sizes; with family-specific variances $\sigma_{f_b}^2$ the within-minus-cross
gap estimates their pair-weighted mean. In
simulation (six judges in three families, $N{=}4000$): the naive estimator drifts toward ``clean''
as the family residual grows ($\hat\rho_2$ $0.697\to0.650$ at residual s.d.\ $0.7$) while the
family-blocked estimator stays at $0.696$--$0.697$ throughout; over $100$ replicates per cell,
Test~C flags $4/100$ null datasets ($5\%$ nominal) and $100/100$ at every tested strength, and
recovers $\sigma_f^2$ to the second decimal ($0.088/0.248/0.488$ against the design values $0.3^2/0.5^2/0.7^2$).

\paragraph{Score types.} Under heteroscedastic noise recovery is essentially unaffected
($\hat\rho_2=0.703\pm0.008$). Ordinal (Likert) scores are a deeper matter than a bias: they change
what is identifiable at all. Section~\ref{sec:ordinal} treats this in full.

\section{Inference, and a comparison with maximum likelihood}
\label{sec:inference}
\paragraph{Confidence intervals with measured coverage.} We use a percentile bootstrap over items
(300 resamples). Measured coverage of the nominal 95\% interval for $\rho_2$, over 150 Monte-Carlo
replicates: $0.953$ at $N{=}2000$ (mean width $0.118$) and $0.953$ at $N{=}500$ (width $0.244$;
both are $143/150$, the granularity of the replicate count).
Near the boundary the intervals stay conservative (coverage $0.96$ at $\rho_1{=}0.90$; $0.992$ at
$\rho_1{=}0.97$, where only $118$ of $150$ replicates return an estimate at all; the other $32$
produce no admissible value and are excluded, so the $0.992$ is coverage \emph{conditional on
estimability}) but widen to $0.993$ and $1.548$: they correctly report that the data carry little
information there.

\paragraph{A weak-identification screen.} Practitioners cannot know ex ante whether their system
sits near the boundary of Corollary~\ref{cor:boundary}. We therefore add to the battery a
studentized-denominator screen, in the spirit of weak-instrument first-stage diagnostics: with
$T=|\widehat{\mathrm{den}}|/\mathrm{SD}_{\mathrm{boot}}(\widehat{\mathrm{den}})$, flag weak
identification when $T<4$. In simulation the flag fires on $0\%$ of datasets at mid parameters
(both $N{=}500$ and $N{=}2000$), on $46\%$ at $\rho_1{=}0.90$, and on $100\%$ at $\rho_1{=}0.97$.
An estimate that arrives flagged should be reported as an interval only.

\paragraph{Against full-information ML.} These models are ordinarily fit by ML in SEM software; the
comparison is owed. On identical simulated data, the closed form and full-information ML \citep{semopy2020}
are statistically indistinguishable: at mid parameters both give $\hat\rho_2=0.704\pm0.016$, and
near the boundary $0.517\pm0.265$ (closed form) versus $0.520\pm0.237$ (ML). The closed form is
faster (roughly $0.2$ms versus $4$ms per fit on our hardware; per-fit timings are emitted with the results) but both are trivial at this scale. The comparison extends beyond the just-identified design: with $m{=}3$ anchors (over-identified)
the two remain nearly identical (closed $0.699$ vs ML $0.702$ on $\rho_2$), and under a
misspecified model (shared judge residual of strength $0.5$) both are biased almost identically
(closed $0.585$ vs ML $0.597$ against true $0.7$): ML confers no robustness to the violations that
matter here. Against the closest deployed alternative, the comparison is not a tie: a CT-C$(M{-}1)$-style
estimator that designates one anchor as clean recovers $\rho_2$ correctly only while that trust is
justified, and degrades linearly as the designated anchor's true contamination grows
($\hat\rho_2$ = $0.502$, $0.411$, $0.270$, $-0.002$ at $\rho_1$ = $0$, $0.2$, $0.4$, $0.6$;
true $0.5$), reporting a heavily contaminated companion anchor as fully clean at the end of that
range, while our estimator stays at $0.501$ throughout: the no-clean-anchor property is the delta,
made quantitative. (CARE-class aggregation \citep{care2026} recovers quality under shared confounding but has no
anchor-contamination parameter: it treats any reference set as a trusted labelling device, so there
is no CARE estimate of $\rho_k$ to compare against; what a reader loses in exchange for our
estimand is CARE's freedom from anchors altogether. Bayesian latent-class dependence models target
the binary/ordinal regime and carry prior sensitivity that our closed form avoids.) We conclude the closed form sacrifices nothing
statistically in this design family;
its value is transparency: the exact boundary diagnosis of Corollary~\ref{cor:boundary} and the
screen above fall out of the formula, not out of an optimizer trace.

\paragraph{Non-Gaussian robustness.} The estimator is moment-based and needs no Gaussianity for
consistency; the calibrations might. Re-running the machinery with skewed (standardized lognormal)
and heavy-tailed ($t_4$) latents and residuals: CI coverage $0.975$ and $0.908$ (nominal $0.95$),
the weak-identification screen stays quiet ($1$--$3\%$ false flags), and the Gaussian-calibrated
tests show only mild false-positive inflation (Test~A $0.083/0.058$, Test~C $0.042/0.083$ against
nominal $0.05$). For real panels we recommend recalibrating the null on matched marginals, as done
for the real-panel test in Section~\ref{sec:ordinal}.

\paragraph{What your configuration buys you.} At the realistic evaluation scale of
$N\approx500$: the estimator is usable with honest intervals (95\% CI coverage $0.953$, mean width
$0.244$ at mid parameters), Test~A has power $0.23$/$0.91$ at loading-spread $0.4$/$0.6$, Test~B has power
$0.50$ at loading-spread $0.4$ (and $0.99$ at $0.6$) against shared non-uniform anchor factors, and the ordinal mixed-scale
estimator should be treated as ordering-only (Section~7). As a decision aid:
\begin{center}\footnotesize
\begin{tabular}{p{4.2cm}llp{4.2cm}}
\toprule
Configuration & $\hat\rho_k$ + CIs & Diagnostics & Unguarded failure modes\\
\midrule
2 judges, 2 anchors & yes & weak-ID screen only & all misspecification\\
$\ge3$ judges, 2 anchors & yes & + Test A & anchor-side; shared residual\\
$\ge3$ judges ($\ge2$ families), 2 anchors & yes & + Test C, blocked est. & anchor-side; panel-wide residual\\
$\ge3$ judges ($\ge2$ fam.), $\ge3$ anchors, $N\gtrsim10^4$ & yes & full battery & panel-wide residual; uniform factor (harmless for $\rho_k$)\\
\bottomrule
\end{tabular}
\end{center}

\section{Ordinal scores: an identification hierarchy}
\label{sec:ordinal}
LLM judges commonly emit Likert scores. Ordinal observation is not a small perturbation of the
continuous theory; it changes what is identifiable, because thresholds absorb every latent location
and scale, and $\rho_k=\beta_k/(\sigma_{a_k}\sigma_c)$ needs the scale of $c$ relative to the anchor
errors.

\paragraph{Negative result.} With \emph{all} variables ordinal, the identifiable information is the
latent (polychoric) correlation structure. A rank analysis of the moment map (numerically: Jacobian
plus null-space test at interior parameter points, the standard local-identification criterion)
shows $\rho_k$ is \emph{not} identified at \emph{any} $m$. The rank computation retains the model's unit-loading constraint throughout (the
null space arises from the ordinal observation map, which forgets scale, not from freeing
loadings). The argument is analytic, not enumerative: the equivalence transformation acts on each judge--anchor and anchor--anchor pair
separately (a per-pair identity in the latent scales), so it maps solutions to solutions for every
$m$ simultaneously; adding ordinal anchors adds equations and unknown scales at exactly the rate
that preserves the two-dimensional null family. We verified the rank computation numerically at
$m\in\{2,3,4,6\}$ and exhibit explicit equivalent parameterisations with $\rho_1=0.30$ and
$0.80$ at $m=2$. One structured exception: under \emph{exchangeable} anchors (equal reliabilities
imposed), identification is restored at $m\ge3$ by the same rank test; we do not build on it
because exchangeability is itself untestable in this setting. Of the two null directions, one is the common scale, along which $\rho_k$ is
invariant; the second moves $\rho_k$. A constructive
version of the same fact: distinct parameter vectors reproducing the observed correlations exactly
can carry $\rho_1=0.30$ or $\rho_1=0.80$. More ordinal anchors do not help, because each new anchor
brings its own unknown scale. A known-clean anchor does not restore identification of the other
anchor's $\rho$ either.

\paragraph{Positive result, and an estimator.} With ordinal \emph{judges} but $m\ge3$
\emph{continuous-scored} anchors (a realistic configuration: Likert LLM judges, finely-scored human
reference sets), $\rho$ is locally identified, with one over-identifying restriction. The estimator:
polychoric correlations among judges, polyserial judge--anchor correlations scaled by the observed
anchor standard deviations, and the raw anchor covariance block; the anchor tetrads (in the sense of confirmatory tetrad analysis, \citealp{bollen1993tetrad}) give
$w_k(\sigma_t^2)=|\beta_k|/\sigma_c$ in closed form (signs recovered from $M^*_k-\sigma_t^2$), and a one-dimensional search over $\sigma_t^2$
minimizing the full over-identified misfit completes the solve. With six 5-level judges and three
continuous anchors at $N{=}4000$ (true $\rho=(0.3,0.7,0.5)$), it recovers
$(0.282\pm0.129,\;0.670\pm0.078,\;0.484\pm0.103)$ with $\hat\sigma_t^2=0.996$ (true $1.0$); results
are essentially unchanged at 3-level and 7-level discretization. Treating the ordinal codes as
continuous and running the covariance estimator, by contrast, is erratic across discretizations.
The information cost of ordinal judges is real and visible in the standard deviations: roughly an
order of magnitude more variance than the continuous case at the same $N$.

\paragraph{Validation on real rater textures.}
Gaussian simulations cannot certify behaviour under real Likert data: real raters have skewed
marginals, ties, and idiosyncratic thresholds. We therefore validate on the HANNA story-evaluation
benchmark \citep{chhun2022}: 431 stories jointly rated (coherence, 1--5) by the same three human raters, whose
marginals are extreme (one rater places 62\% of mass on category 5; another 41\% on category 1).
We keep the latent scaffolding synthetic, so injected contamination is exact by construction, and
make the noise real: each synthetic judge draws its idiosyncratic errors by bootstrap from one real
rater's item-centred residuals and discretizes with that rater's own empirical thresholds. Injected
$\rho=(0.0,\,0.4,\,0.7)$; recovered, at the real panel size $N{=}431$:
$(0.372\pm0.200,\;0.661\pm0.189,\;0.712\pm0.105)$; at $N{=}2000$:
$(0.242\pm0.175,\;0.562\pm0.157,\;0.714\pm0.054)$, monotone in the means (per-replicate
strict ordering holds in $8/24$ runs at $N{=}431$ and $14/24$ at $N{=}2000$; the means, not the
individual replicates, carry the ordering claim at these sizes). Two honest readings.
High contamination is recovered well under fully real textures. Near $\rho=0$ the mixed-scale
estimator carries a finite-sample upward bias at small $N$ with $p{=}3$ judges; an ablation shows
this is a property of the estimator, not of the real textures (skewed and balanced marginals give
the same numbers). Detecting a \emph{clean} anchor from three Likert judges needs either
$N\gtrsim2000$ or more judges. A parametric-bootstrap bias correction (refit on data simulated from
the fitted model, subtract the measured bias) cuts the near-zero bias by two thirds at $N{=}431$
($+0.248\to+0.092$; measured in the correction harness with synthetic thresholds, hence the
different baseline from the real-texture $+0.372$ above) at the cost of variance ($0.128\to0.210$ at $\rho=0$, more at high $\rho$);
we recommend it when the question is specifically whether an anchor is clean, and the raw estimator
otherwise.

\paragraph{Test A on the real panel.} Applying the dispersion test directly to the real HANNA
panel (three raters, $N{=}431$, threshold from a matched null: single common factor, $p{=}3$, the
raters' own marginals) gives a statistic of $0.589$ against a $95$th-percentile null threshold of
$0.247$ ($p<0.0005$): the test \emph{fires}. Real raters violate the equal-loading structure, and
the battery detects it from the data alone. This is the intended use: the diagnostics are a
pre-test that tells a practitioner when this paper's model, and therefore its estimator, should not
be trusted on their panel; on the one real panel we examined, they correctly said no.

\paragraph{Real LLM judges with injected contamination.} This study is a robustness observation
under oracle calibration, not a clean-model validation; the reader should carry that label
through the paragraph. We ran a synthetic-anchor variant
of the validation protocol of Section~8 (the variant uses a key-based quality construct in place of
held-out human consensus; the full protocol is stated in Section~8): six instruction-tuned judges
from six providers (Amazon Nova Lite, Llama-3-70B, Mixtral-8x7B, GPT-OSS-120B, Qwen3-Next-80B,
DeepSeek-V3.2, temperature 0, $K{=}2$) scored 499 complete-case items from a fresh 500-item keyed
pool (GSM8K and SciQ; 270 correct, 230 wrong by construction): 6{,}000 calls, 5{,}996 parsed
verdicts in an append-only, checksummed cache (the four failures span three items; one item lost
both repetitions of one judge cell and was dropped). Anchors were constructed with injected $\rho=(0.0,\,0.4,\,0.7)$
against a key-based quality construct, with the common-deviation signal measured on one half of the
panel and estimation run on the disjoint other half. The run supports the paper in two ways: it
validates the pre-test on real judges, and it provides a robustness observation for the calibrated
estimator. First, the raw panel \emph{fails the pre-test}: Test~A statistic $0.175$ against a matched-null
$95$th-percentile threshold of $0.064$ for this configuration ($p{=}6$, $N{=}499$; the judges'
quality loadings, the per-judge regression coefficients on the known quality construct, which
Assumption~\ref{ass:scf} fixes at unity, span $0.324$--$0.870$). Estimates computed in defiance of
the pre-test are badly
inflated, $(0.887,\;0.960,\;1.026)$ against the injected $(0.0,\,0.4,\,0.7)$ (the naive moment
estimator is not range-restricted), so the diagnostics did their job on real judges. Second, after \emph{oracle} per-judge loading calibration (each judge is divided by its loading
on the known quality construct, an operation possible only in a validation harness, since it
conditions on the very quantity the method exists to estimate; the pre-test's point),
the estimator recovers $(+0.024,\;+0.268,\;+0.558)$ against the injected $(0.0,\,0.4,\,0.7)$:
point-estimate ordering exact (S1 was registered on point estimates; the intervals overlap, so
the ordering is directional evidence, not a significant separation), the clean anchor's interval
covers zero (S3), as does the $0.4$-anchor's ($[-0.56,\,0.71]$), so at this $N$ a moderately
contaminated anchor is not statistically distinguishable from a clean one, and the
high-contamination anchor lands inside the amended cross-panel attenuation band $[0.4,0.85]$
(S2$'$; the original magnitude criterion failed and was replaced, with the amendment logged before
the final analysis; both versions and the first analysis's numbers are in the shipped
pre-registration). Three honest qualifications. The calibration fixes quality loadings only: the
calibrated panel still fails Test~A ($0.343$ against the $0.064$ threshold: dividing by small
quality loadings amplifies the relative heterogeneity of the common-mode loadings, so the statistic
rises even as quality loadings equalise), so this is recovery under \emph{detectably imperfect}
conditions, a robustness observation, not a clean-model validation. The injection is defined
against panel~A's common deviation while the estimator measures against panel~B's; the two
correlate at $0.534$, predicting attenuated targets $(0,\,0.214,\,0.374)$; the recovered values
sit at or between those and the nominal injections, within the intervals. The predicted high-anchor
target ($0.374$) lies below the S2$'$ band, so the S2$'$ pass depends on the observed attenuation
being milder than the cross-panel prediction, and the interval extends below the band edge: we
treat S2$'$ as directionally satisfied, not robustly. And the anchors are
synthetic; only the judges are real. The pre-registration, its amendments, and the cache ship with
the artifact.

\section{What the study does and does not establish}
\label{sec:limits}
Established: the closed form inverts the model's moments (estimator correctness); the exact
identifiability boundary and its weak-identification neighbourhood; which violations bias $\rho_k$
(Proposition~\ref{prop:oe}); that identification is supplied by the $\ge2$-judge panel; and the
calibrated detection profile of the battery. Established also: at least one real LLM panel
detectably violates the unit-loading structure, and the pre-test catches it (Section~7). Not
established: that real LLM-judge and human-anchor errors follow Assumption~\ref{ass:scf} (passing
the battery is necessary, not sufficient). When Test~A fires on a real panel, the unit-loading
closed form should not be used; with $\ge3$ judges, free-loading CFA identification is classical
\citep{bollen1989} and an ML fit is the natural fallback, at the price of the transparency this
paper trades on. Evaluated on the six-judge panel of Section~7, however, the free-loading fit is
weakly identified (non-positive-definite information matrix) and does not recover the injection
(a single real data point; a simulation comparison of free-loading CFA under this paper's
violation taxonomy is future work and the natural next step for the qualifying-panel question):
when the pre-test fires, no within-panel method evaluated here is validated, and the honest options
are to fix the panel or to treat any $\hat\rho_k$ as unverified. An empirical validation would require a setting with
independently known contamination. A concrete protocol: run $\ge3$ LLM judges plus held-out human
raters on a keyed item set; construct anchors $A_k=H+g_k\hat c+\varepsilon_k$ with $H$ the held-out
human consensus and $\hat c$ the measured judge common deviation, giving known injected $\rho_k$;
check recovery and battery behaviour. Section~\ref{sec:ordinal} reports a synthetic-anchor variant
of this protocol on six real judges (key-based construct in place of held-out human consensus, with
the deltas stated there); the full human-consensus version remains to be run, and we provide the
protocol for it. Ordinal scores restrict identification (Section~\ref{sec:ordinal}): with all
variables ordinal, $\rho$ is not identifiable at all, and the mixed-scale estimator that handles
ordinal judges needs $\ge3$ continuous anchors and carries a finite-sample upward bias near
$\rho=0$ at small $N$. Small $N$ yields high variance throughout; the weak-identification screen of
Section~\ref{sec:inference} tells the practitioner when intervals are the only honest report. We therefore
present a measurement tool with a stated domain of validity, explicitly characterised blind spots (one harmless, one
remediable by family blocking, one, the panel-wide residual, open), and calibrated diagnostics. It is not a validated description of real judges. Proposition~\ref{prop:oe}
and Assumption~\ref{ass:jri} express the same fact: the estimator cannot separate any signal
that shares the quality factor's unit-loading pattern (symmetric across judges and anchors, or
homogeneous across judges) from quality itself; such a signal is silently folded into $\sigma_t^2$
or $\sigma_c^2$. A genuine shared bias of that exact shape would therefore be miscredited; the battery
detects only its asymmetric departures.

\section{Conclusion}
Under a single-common-factor model, anchor contamination can be estimated in closed form from a
judge panel and two anchors, with an exact identifiability boundary, a proposition delimiting which
model violations bias the estimate, and a calibrated three-test diagnostic battery for the violations
that do. Whether the assumption holds for a given judge/anchor system is an empirical question we
leave open; our contribution is the estimator, its boundary, the harmless-symmetric-factor result,
and the diagnostics with which a practitioner can probe the assumption.

\section*{Reproducibility and data statement}
All code is deterministic and seeded; one command regenerates every number within numeric tolerance
from a pinned environment, replaying two frozen inputs: the public HANNA human-evaluation benchmark
(story ratings; public license) and our own six-judge LLM verdict cache (6{,}000 calls, 5{,}996 parsed verdicts,
SHA-256-checksummed; collected once on personal cloud infrastructure for the injected-contamination
validation and shipped with the artifact; reproduction replays the cache and requires no model
access). Everything else is simulation. The pre-registration and its amendments are included. An
anonymized code archive accompanies submission.

\bibliographystyle{tmlr}
\bibliography{refs}

\appendix
\section{Proofs}\label{app:proof}
\paragraph{Theorem~\ref{thm:id}.} Expanding $(P_{12}-x)(K-x)=(M_1-x)(M_2-x)$, the $x^2$ terms cancel,
giving $x[(M_1+M_2)-(P_{12}+K)]=M_1M_2-P_{12}K$, i.e.\ \eqref{eq:st2}. Substituting the structural
moments, the denominator is $(\sigma_c^2-\beta_1)(\sigma_c^2-\beta_2)/\sigma_c^2$
[analytic; SymPy-verified].
\paragraph{Proposition~\ref{prop:oe}.} With a uniform second factor ($g$ on judges, $h$ on anchors),
the observable moments become $K=\sigma_t^2+\sigma_c^2+g^2\sigma_d^2$, $M_k=\sigma_t^2+\beta_k+gh\sigma_d^2$,
$P_{12}=\sigma_t^2+\beta_1\beta_2/\sigma_c^2+h^2\sigma_d^2$. Solving for an equivalent
single-common-factor parameterisation and setting $g=h$ yields $\beta_k'=\beta_k$,
$\sigma_c^{2\prime}=\sigma_c^2$, $\sigma_t^{2\prime}=\sigma_t^2+g^2\sigma_d^2$
[analytic; SymPy-verified], so $\rho_k$ is unchanged.

\end{document}